\documentstyle[aps,epsf,multicol]{revtex}
\begin{document}
\def\I.#1{\it #1}
\def\B.#1{{\bbox#1}}
\def\C.#1{{\cal  #1}}

\title{Thermodynamic Formalism of the Harmonic Measure of
Diffusion Limited Aggregates: Phase Transition and Converged
$f(\alpha)$}
\author { Mogens H. Jensen$^*$, Anders Levermann$^{**}$, Joachim
Mathiesen$^*$, Benny Davidovitch$^{**}$ and Itamar Procaccia$^{**}$}
\address{$^*$ The Niels Bohr Institute, Blegdamsvej, Copenhagen,
Denmark\\
$^{**}$Department of~~Chemical Physics, The Weizmann Institute of
Science, Rehovot 76100, Israel}
\maketitle


%
%

\begin{abstract}
We study the nature of the phase transition in the multifractal
formalism of
the harmonic measure of Diffusion Limited Aggregates (DLA). Contrary to
previous work that
relied on random walk simulations or ad-hoc models to estimate the low
probability events of deep fjord
penetration, we employ the method of iterated conformal
maps to  obtain an accurate computation of the probability of the
rarest events. We resolve probabilities as small as $10^{-70}$. We show
that the generalized dimensions
$D_q$ are infinite for $q<q^*$, where $q^*= -0.17\pm 0.02$. In the
language of $f(\alpha)$ this means that
$\alpha_{\rm max}$ is finite. We present a converged $f(\alpha)$ curve.
\end{abstract}
\begin{multicols}{2}
Since its introduction in 1981 \cite{81WS} the model of Diffusion
Limited Aggregation (DLA) has
posed a challenge to our understanding of fractal and multifractal
phenomena.
DLA is a paradigmatic example for the spontaneous generation of fractal
objects by simple dynamical rules  (being generated by random walkers);
its harmonic measure,
which is the probability for a random walker to hit the surface, had
been one of the
first studied examples of multifractal measures outside the realm of
ergodic measures
in dynamical systems \cite{86HMP}. The multifractal properties stem from
the extreme contrast between
the probability to hit the tips of the DLA compared with penetrating the
fjords.

The multifractal properties of the harmonic measure of the DLA are
conveniently
studied in the context of the generalized dimensions $D_q$, and the
associated
$f(\alpha)$ function \cite{83HP,86HJKSP}. The simplest definition of the
generalized dimensions
is in terms of a uniform covering of the boundary of a DLA cluster
with boxes of size $\ell$, and measuring the probability for a random
walker coming from infinity to hit a piece of boundary which belongs to
the $i$'th box.
Denoting this probability by $P_i(\ell)$, one considers \cite{83HP}
\begin{equation}
D_q \equiv \lim_{\ell \to 0}\frac{1}{q-1}\frac{\log\sum_i
P_i^q(\ell)}{\log\ell} \ .
\end{equation}
It is well known by now that the existence of an interesting spectrum of
values $D_q$ is
related to the probabilities $P_i(\ell)$ having a spectrum of
``singularities" in the
sense that $P_i(\ell) \sim \ell^\alpha$
with $\alpha$ taking on values from a range $\alpha_{\rm min}\le
\alpha\le \alpha_{\rm max}$.
The frequency of observation of a particular value of $\alpha$ is
determined by the the function
$f(\alpha)$ where (with $\tau(q)\equiv (q-1)D_q$)
\begin{equation}
f(\alpha) = \alpha q(\alpha)-\tau\Big(q\left(\alpha\right)\Big)\ ,\quad
\frac{\partial \tau(q)}{\partial q}=\alpha(q) \ .
\end{equation}

The understanding of the multifractal properties and the associated
$f(\alpha)$ spectrum of DLA clusters have been a long standing
issue. Of particular interest are the values of the minimal
and maximal values, $\alpha_{min}$ and $\alpha_{max}$, relating
to the largest and smallest growth probabilities, respectively.
As a DLA cluster grows the large branches screen the deep fjords
more and more and the probability for a random walker to get into
these fjords (say around the seed of the cluster) becomes
smaller and smaller. A small growth probability corresponds  to a
large value of $\alpha$.
Previous literature hardly agrees about the actual value of
$\alpha_{max}$.
Ensemble averages of the harmonic measure
of DLA clusters indicated a rather large value of $\alpha_{max} \sim 8$
\cite{averages}.
In subsequent experiments on non-Newtonian fluids \cite{Nittmann}
and on viscous fingers \cite{viscous},
similar large values of $\alpha_{max}$ were also observed.
These numerical and experimental indications of a very large value
of $\alpha_{max}$ led to a conjecture that, in the limit
of a large, self-similar cluster some fjords will be exponentially
screened and thus causing $\alpha_{max} \to \infty$ \cite{Bohr}.

If indeed $\alpha_{max} \to \infty$, this can be interpreted
as a phase transition \cite{Predrag} (non-analyticity) in the $q$
dependence
of $D_q$, at a value of $q$ satisfying $q\ge 0$. If the transition takes
place for a value
$q < 0$ then $\alpha_{max}$ is finite. Lee and Stanley \cite{88LS}
proposed that $\alpha_{max}$
diverges like $R^2/\ln{R}$ with $R$ being the radius of the cluster.
Schwarzer et al. \cite{schwarzer} proposed that
$\alpha_{max}$ diverges only logarithmically in the number of added
particles.
Blumenfeld and Aharony \cite{amnon} proposed that channel-shaped fjords
are important and proposed that $\alpha_{max} \sim {{M^x} \over {ln~M}}$
where $M$ is the mass of the cluster;
Harris and Cohen \cite{90HC}, on the other hand, argued that straight
channels might be so
rare that they do not make a noticeable contribution, and
$\alpha_{max}$ is finite, in agreement with Ball and Blumenfeld who
proposed \cite{Ball}
that $\alpha_{max}$ is bounded below 11. Obviously, the issue was not
quite settled. The difficulty
is that it is very hard to estimate the smallest growth probabilities
using models or direct numerical simulations.

In this Letter we use the method of iterated conformal maps to offer an
accurate
determination of the probability for the rarest events. We propose that
using this method we can settle the issue in a conclusive way. Our
result
is that $\alpha_{max}$ exists and the phase transition occurs at a $q$
value
that is slightly negative. In this method one
studies DLA by constructing
$\Phi^{(n)}(w)$ which conformally maps the exterior of the unit circle
$e^{i\theta}$ in the
mathematical $w$--plane onto the complement of the (simply-connected)
cluster of $n$ particles in the physical $z$--plane
\cite{98HL,99DHOPSS,00DFHP}.
The unit circle is
mapped onto the boundary of the cluster. The map $\Phi^{(n)}(w)$ is
made from compositions of elementary maps $\phi_{\lambda,\theta}$,
\begin{equation}
\Phi^{(n)}(w) = \Phi^{(n-1)}(\phi_{\lambda_{n},\theta_{n}}(w)) \ ,
\label{recurs}
\end{equation}
where the elementary map $\phi_{\lambda,\theta}$ transforms the unit
circle to a circle with a semi-circular ``bump" of linear size $\sqrt{\lambda}$ around
the point $w=e^{i\theta}$. We use below the same map $\phi_{\lambda,\theta}$ that
was employed in \cite{98HL,99DHOPSS,00DFHP,00DP,00DLP}.
With this map $\Phi^{(n)}(w)$ adds on a semi-circular
new bump to the image of the unit circle under $\Phi^{(n-1)}(w)$. The
bumps in the $z$-plane simulate the accreted particles in
the physical space formulation of the growth process. Since we want
to have {\em fixed size} bumps in the physical space, say
of fixed area $\lambda_0$, we choose in the $n$th step
\begin{equation}
\lambda_{n} = \frac{\lambda_0}{|{\Phi^{(n-1)}}' (e^{i \theta_n})|^2} \ .
\label{lambdan}
\end{equation}
The recursive dynamics can be represented as iterations
of the map $\phi_{\lambda_{n},\theta_{n}}(w)$,
\begin{equation}
\Phi^{(n)}(w) =
\phi_{\lambda_1,\theta_{1}}\circ\phi_{\lambda_2,\theta_{2}}\circ\dots\circ
\phi_{\lambda_n,\theta_{n}}(\omega)\ . \label{comp}
\end{equation}
It had been demonstrated before that this method represents DLA
accurately, providing many analytic insights that are not available
otherwise \cite{00DP,00DLP}. For our purposes here we quote a result
established in \cite{99DHOPSS},
which is
\begin{equation}
\langle \lambda^q_n \rangle\equiv
(1/2\pi)\int_0^{2\pi}\lambda^q_n(\theta) d\theta
 \sim n^{-2qD_{2q+1}/D} \ . \label{lamDq}
\end{equation}
To compute $\tau(q)$ we rewrite this average as
\begin{equation}
\label{mean}
\langle \lambda^q_n \rangle=
\int ds  \left|\frac{d\theta}{ds}\right| ~\lambda^q_n(s)=\int ds
\frac{\lambda^{q+1/2}(s)}{\sqrt{\lambda_0}} \ ,
\end{equation}
where $s$ is the arc-length of the physical boundary of the cluster. In
the last equality we used the fact that
$|d\theta/ds|=\sqrt{\lambda_n/\lambda_0}$. We stress at this point that
in order to measure
these moments for $q\le 0$ we {\em must} go into arc-length representation.

To make this crucial point clear
we discuss briefly what happens if one attempts to compute the moments
from the definition
(\ref{lamDq}). Having at hand the conformal map
$\Phi^{(n)}(e^{i\theta})$,  one
can choose randomly as many points on the unit circle $[0,2\pi]$ as one
wishes,
obtain as many  (accurate) values of $\lambda_n$, and try to compute the
integral
as a finite sum. The problem is of course that using such an approach
{\em the
fjords are not resolved}. To see this we show in Fig.1 panel a the
region
of a typical cluster of 50 000 particles that is being  visited by a
random
search on the unit circle. Like in direct simulations using
random walks, the rarest events are not probed, and no
serious conclusion regarding the phase transition is possible.

\begin{figure}
\epsfxsize=6truecm
\epsfbox{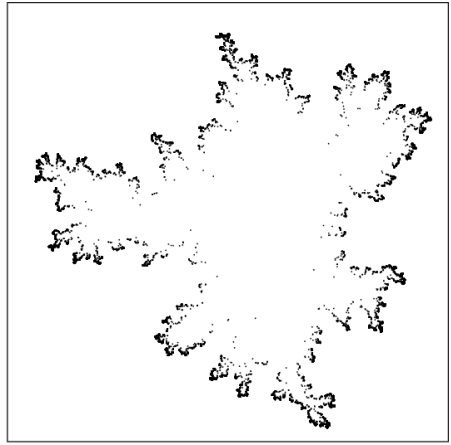}
\vskip 0.5truecm
\epsfxsize=6truecm
\epsfbox{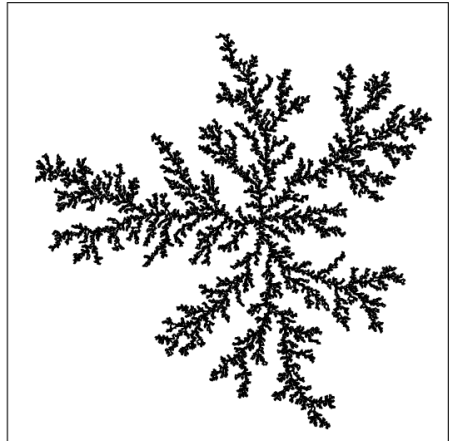}
\caption{panel a: the boundary of the cluster probed by a random search
with respect to the harmonic measure. Panel b: the boundary of the
cluster probed by the
present method.}
\label{patterns}
\end{figure}
Another method that cannot work is to try to compute by sampling on the
arc-length in a naive way. The reason is that the inverse
map $[\Phi^{(n)}]^{-1}(s)$ cannot resolve $\theta$ values that belong
to deep fjords. As the growth proceeds, reparametrization squeezes
the $\theta$ values that map to fjords
into minute intervals, below the computer numerical resolution.
To compute the values of $\lambda_n(s)$ effectively we must use the
full power of our iterated conformal dynamics, carrying the history
with us, to iterate forward and backward at
will to resolve accurately the $\theta$, $\lambda$ values of any given
particle on
the fully grown cluster.

To do this we recognize that every time we grow a semi-circular bump
we generate two new branch-cuts in the map  $\Phi^{(n)}$. We find the
position
on the boundary between every two branch-cuts, and there compute
the value of $\lambda_n$. The first step in our algorithm is to generate
the location
of these points intermediate to the branch-cuts \cite{01BDP}. Each branch-cut has a
preimage on the
unit circle  which will be indexed with 3
indices, $w^{k(\ell)}_{j,\ell}\equiv \exp[i\theta^{k(\ell)}_{j,\ell}]$.
The index $j$ represents the generation when
the branch-cut was created (i.e. when the $j$th particle was grown).
The index $\ell$ stands for the generation at which the analysis is
being done (i.e. when the cluster has $\ell$ particles). The index $k$
represents the position of the
branch-cut along the arc-length, and it is a function of the
generation $\ell$.  Note that
since bumps may overlap during growth, branch-cuts are then
covered, and therefore the maximal $k$, $k_{max} \le 2\ell$.
After each iteration the preimage of each branch-cut moves on
the unit circle, but its physical position remains. This leads
to the equation
that relates the indices of a still exposed branch-cut that was
created at generation $j$ to a later generation $n$:
\begin{eqnarray}
\Phi^{(n)}(w^{k(n)}_{j,n})&\equiv&
\Phi^{(n)}\left(\phi^{-1}_{\lambda_n,\theta_n}
\circ\dots\circ\phi^{-1}_{\lambda_{j+1},\theta_{j+1}}
(w^{\tilde k(j)}_{j,j})\right)\nonumber\\&=& \Phi^{(j)}(w^{\tilde
k(j)}_{j,j}) \ . \label{trick}
\end{eqnarray}
Note that the sorting indices $\tilde k(j)$ are not simply
related to $k(n)$, and needs to be tracked as follows.
Suppose that the list $w^{k(n-1)}_{j,n-1}$ is available. In the $n$th
generation we
choose randomly a new $\theta_n$, and find two new branch-cuts
which on the unit circle are at angles $\theta_n^{\pm}$. If one (or very
rarely more)
branch-cut of the updated list
$\phi^{-1}_{\lambda_n,\theta_n}(w^{k(n-1)}_{j,n-1})$ is covered,
it is eliminated from the
list, and together with the sorted new pair we make the list
$w^{k(n)}_{j,n}$. Having a cluster of $n$ particles we now consider
all neighboring pairs of preimages $w^{k(n)}_{j,n}$ and
$w^{k(n)+1}_{J,n}$,
that very well may have been created at two different generations $j$
and $J$.
The larger of these indices ($J$ without loss of generality)
determines the generation of the intermediate
position at which we want to compute the field. We want to find the
preimage
$u^{k(n)}_{J,J}$ of this mid-point on the unit circle , to compute
$\lambda_{k(n)}$
there accurately. Using definition (\ref{trick}) we find the preimage
\begin{equation}
\arg(u^{k(n)}_{J,J}) = [\arg(w^{\tilde k(J)}_{j,J})+\arg(w^{\tilde
k(J)+1}_{J,J})]/2 \ .
\end{equation}
In Fig.1 panel b we show, for the same cluster of 50 000, the map
$\Phi^{(J)}(u^{k(n)}_{J,J})$
with $k(n)$ running between 1 and $k_{\rm max}$, with
$J$ being the corresponding generation of creation of the mid point. We
see that
now all the particles are probed, and
every single value of $\lambda_{k(n)}$ can be computed. However, to compute
these $\lambda_{k(n)}$
accurately, we define (in analogy to Eq.(\ref{trick})) for every $J<m\le n$
\begin{equation}
u^{k(n)}_{J,m}\equiv \phi^{-1}_{\lambda_m,\theta_m}\circ \dots
\circ
\phi^{-1}_{\lambda_{j+1},\theta_{j+1}}(u^{k(n)}_{J,J}) \ .
\end{equation}
Finally $\lambda_{k(n)}$ is computed from the definition
(\ref{lambdan})
with
\begin{eqnarray}
{\Phi^{(n)}}'(u^{k(n)}_{J,n})
&=&\phi'_{\lambda_n,\theta_n}(u^{k(n)}_{J,n})
\cdots\phi'_{\lambda_{J+1},\theta_{J+1}}(u^{k(n)}_{J,J+1})\nonumber\\&\times&
{\Phi^{(J)}}'(u^{k(n)}_{J,J}) \ . \label{prod}
\end{eqnarray}
This calculation is optimally accurate since we avoid as much
as possible the effects of the exponential shrinking of low probability 
regions on the unit circle.
Each derivative in (\ref{prod}) is computed using information from
a generation in which points on the unit circle are optimally resolved.

\begin{figure}
\epsfxsize=7truecm
\epsfbox{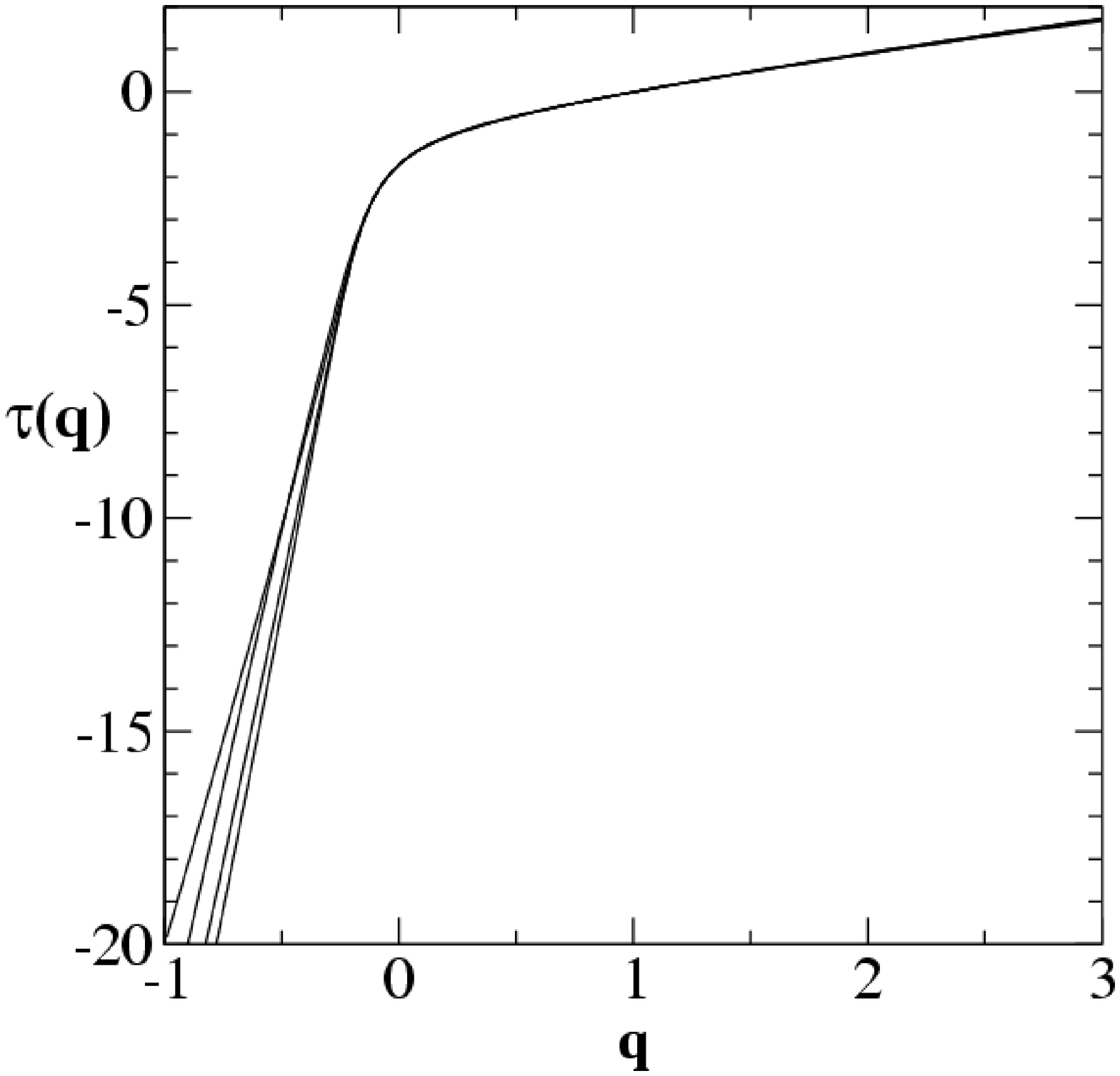}
\vskip 0.5truecm
\epsfxsize=7truecm
\epsfbox{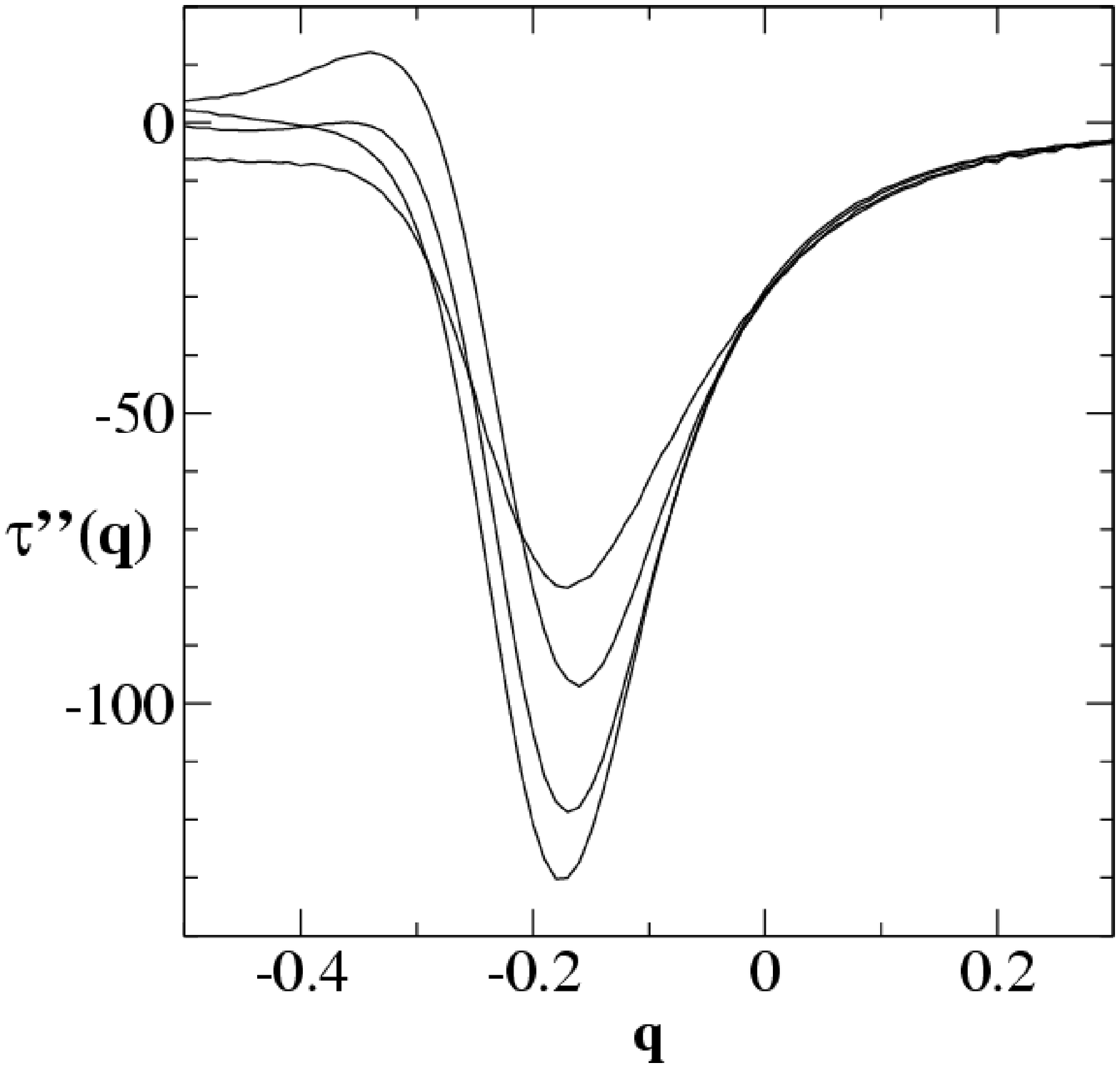}
\caption{Panel a: the calculated function $\tau(q)$ for clusters of $n$
particles,
with $n=$10 000,15 000, 25 000 and 30 000. Panel b: the second
derivative of $\tau(q)$
with respect to $q$.}
\label{tau}
\end{figure}
The integral (\ref{mean}) is then estimated as the finite sum
$\sqrt{\lambda_0}\sum_{k(n)}  \lambda_{k(n)}^q$.
We should stress that for clusters of the order of 30 000 particles we
already
compute, using this algorithm, $\lambda_{k(n)}$ values of the order of
$10^{-70}$.
To find the equivalent small probabilities using random walks would
require
about $10^{70}$ attempts to see them just once. This is of course
impossible,
explaining the lasting confusion about the issue of the phase transition
in
this problem. This also means that all the $f(\alpha)$ curves that were
computed before \cite{averages,87HSM} did not converge. Note that in
our calculation the small values of $\lambda_{k(n)}$ are obtained from
multiplications
rather than additions, and therefore can be trusted.

Having the accurate values $\lambda_{k(n)}$ we can now compute the moments
(\ref{lamDq}).
Since the scaling form on the RHS includes unknown coefficients, we
compute
the values of $\tau(q)$ by dividing $\langle\lambda_n^q\rangle$
by $\langle\lambda_{\bar n}^q\rangle$, estimating
\begin{equation}
\tau(q) \approx -D \frac{\log{\langle\lambda_n^q\rangle}
-\log{\langle\lambda_{\bar n}^q\rangle}}{\log{n}-\log{\bar n}}
\end{equation}
Results for $\tau(q)$ for increasing values of $n$ and $\bar n$ are
shown in
Fig. 2, panel (a). It is seen that the value of $\tau(q)$ appears to
grow
without bound for $q$ negative. The existence of a phase transition is
however best indicated by measuring the derivatives of $\tau(q)$ with
respect
to $q$. In Fig. 2 panel b we show the second derivative, indicating a
phase
transition at a value of $q$ that recedes {\em away} from $q=0$ when $n$
increases. Due to the great accuracy of our measurement of $\lambda$ we
can
estimate already with clusters as small as 20-30 000 the $q$ value of
the phase transition to $q=-0.17\pm 0.02$. The fact that this value
is very close to the converged value can be seen from the $f(\alpha)$
curve which is plotted in Fig. 3. A test of convergence is that the
slope of this function where it becomes essentially linear must agree
with the $q$ value of the phase transition. The straight line shown
in Fig.3 has the slope of -0.17, and it indeed approximates very
accurately the slope of the $f(\alpha)$ curve where it ends and stops
being analytic. The reader should also note that the peak of the
curve agrees with $D\approx 1.71$, as well as the fact that
$\tau(3)$ is also $D$ as expected in this problem.  The value
of $\alpha_{\rm max}$ is close to 20, which is higher than
anything predicted before. It is nevertheless finite. We believe
that this function is well converged, in contradistinction with
past calculations.

\begin{figure}
\epsfxsize=7truecm
\epsfbox{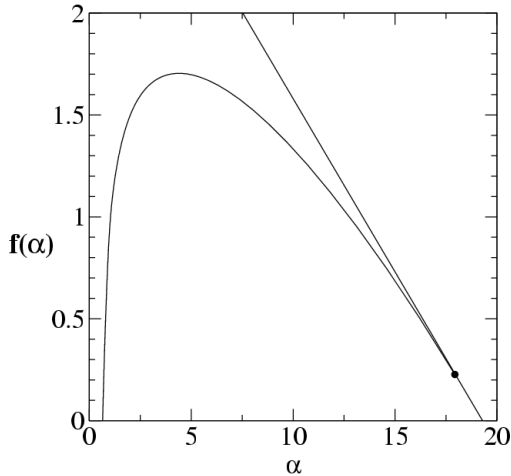}
\caption{The calculated function $f(\alpha)$ using $\tau(q)$ calculated
from a cluster with  $n=30~000$ particles. This $f(\alpha)$ is
almost indistinguishable from the one computed with
$n=25 000$ particles. We propose that this
function is well converged. The black dot denotes where the
curve ends, being tangent to the line with slope -0.17.}
\label{tau}
\end{figure}

%
%

\acknowledgments
This work has been supported in part by the
Petroleum Research Fund, The  European Commission under the
TMR program and the Naftali and Anna
Backenroth-Bronicki Fund for Research in Chaos and Complexity. A. L.
supported by a fellowship of the Minerva Foundation, Munich, Germany.


%
%

\end{multicols}

\begin{thebibliography}{1}

\bibitem{81WS}
T.A. Witten and L. Sander, Phys. Rev. Lett. {\bf 47}, 1400 (1981).

\bibitem{86HMP}
T.C. Halsey, P. Meakin and I. Procaccia, Phys. Rev. Lett. {\bf 56}, 854
(1986).

\bibitem{83HP}
H.G.E. Hentschel and I. Procaccia, Physica  D{\bf 8}, 435 (1983).

\bibitem{86HJKSP}
T.C. Halsey, M.H. Jensen, L.P. Kadanoff, I. Procaccia and B. Shraiman,
            Phys. Rev. A {\bf 33}, 1141 (1986).

\bibitem{averages}
C. Amitrano, A. Coniglio and F. di Liberto, Phys.Rev.Lett.
{\bf 57}, 1098 (1986).

\bibitem{Nittmann}
J. Nittmann, H.E. Stanley, E. Touboul and G. Daccord, Phys.Rev.Lett.
{\bf 58}, 619 (1987).

\bibitem{viscous}
K.J. M\aa l\o y, F. Boger, J. Feder, and T. J\o ssang, in
``Time-Dependent
Effects in Disordered Materials", eds. R. Pynn and T. Riste,
(Plenum, New York, 1987), p.111.

\bibitem{Bohr}
T. Bohr, P. Cvitanovi\'c and M.H. Jensen, Europhys.Lett. {\bf 6},
445 (1988).

\bibitem{Predrag}
            P. Cvitanovi\'c, in Proceedings of ``XIV Colloquium on Group
Theoretical
            Methods in Physics", ed. R.Gilmore (World Scientific,
Singapore 1987);
            in ``Non-Linear Evolution and Chaotic Phenomena", eds. P.
Zweifel, G.
            Gallavotti and M. Anile (Plenum, New York, 1988).

\bibitem{88LS}
J. Lee and H.E. Stanley, Phys.Rev.Lett. {\bf 61}, 2945 (1988).

 \bibitem{schwarzer}
            S. Schwarzer, J. Lee, A. Bunde, S. Havlin, H.E. Roman and
H.E. Stanley,
            Phys.Rev. Lett. {\bf 65}, 603 (1990).

\bibitem{amnon}
            R. Blumenfeld and A. Aharony, Phys.Rev.Lett.
            {\bf 62}, 2977 (1989)

\bibitem{90HC}
A.B. Harris and M. Cohen, Phys.Rev.A {\bf 41}, 971 (1990).


\bibitem{Ball}
R.C. Ball and R. Blumenfeld, Phys. Rev. {\bf A 44}, R828, (1991).


 \bibitem{98HL} M.B. Hastings and L.S. Levitov, Physica D {\bf 116},
244 (1998).

 \bibitem{99DHOPSS} B. Davidovitch, H.G.E. Hentschel, Z. Olami,
   I.Procaccia,
   L.M. Sander, and E. Somfai,
   Phys. Rev. E, {\bf 59} 1368 (1999).

\bibitem{00DFHP}
   B. Davidovitch, M.J. Feigenbaum, H.G.E. Hentschel and I. Procaccia,
   Phys. Rev. E {\bf 62}, 1706 (2000).

\bibitem{00DP}
B. Davidovitch and I. Procaccia, Phys. Rev. Lett., {\bf 85}
       3608-3611 (2000).

\bibitem{00DLP}
B. Davidovitch, A. Levermann, I. Procaccia, Phys. Rev. E, {\bf 62}
R5919.

\bibitem{01BDP}
F. Barra, B. Davidovitch and I. Procaccia, ``Iterated Conformal Dynacmics
and Laplacian Growth", submitted to Phys.Rev. E, also cond-mat/0105608.

\bibitem{87HSM}
Y. Hayakawa, S. Sato and M. Matushita, Phys. Rev. A {\bf 36}, 1963
(1987).



\end{thebibliography}
\end{document}